\documentstyle[12pt]{article}
\def\Re{{\cal R \mskip-4mu \lower.1ex \hbox{\it e}\,}}
\def\Im{{\cal I \mskip-5mu \lower.1ex \hbox{\it m}\,}}
\def\ie{{\it i.e.}}
\def\eg{{\it e.g.}}

\def\etal{{\it et al.}}
\def\ibid{{\it ibid}.}
\def\sub#1{_{\lower.25ex\hbox{$\scriptstyle#1$}}}
\def\sul#1{_{\kern-.1em#1}}
\def\sll#1{_{\kern-.2em#1}}
\def\sbl#1{_{\kern-.1em\lower.25ex\hbox{$\scriptstyle#1$}}}
\def\ssb#1{_{\lower.25ex\hbox{$\scriptscriptstyle#1$}}}
\def\sbb#1{_{\lower.4ex\hbox{$\scriptstyle#1$}}}

\def\to{\rightarrow}
\def\mh{\ifmmode m\sbl H \else $m\sbl H$\fi}
\def\mch{\ifmmode m_{H^\pm} \else $m_{H^\pm}$\fi}
\def\mt{\ifmmode m_t\else $m_t$\fi}
\def\mc{\ifmmode m_c\else $m_c$\fi}
\def\mz{\ifmmode M_Z\else $M_Z$\fi}
\def\mw{\ifmmode M_W\else $M_W$\fi}
\def\mws{\ifmmode M_W^2 \else $M_W^2$\fi}
\def\mhs{\ifmmode m_H^2 \else $m_H^2$\fi}
\def\mzs{\ifmmode M_Z^2 \else $M_Z^2$\fi}
\def\mts{\ifmmode m_t^2 \else $m_t^2$\fi}
\def\om{{\omega}}
\def\mcs{\ifmmode m_c^2 \else $m_c^2$\fi}
\def\mchs{\ifmmode m_{H^\pm}^2 \else $m_{H^\pm}^2$\fi}
\def\ztwo{\ifmmode Z_2\else $Z_2$\fi}
\def\zone{\ifmmode Z_1\else $Z_1$\fi}
\def\mtwo{\ifmmode M_2\else $M_2$\fi}
\def\mone{\ifmmode M_1\else $M_1$\fi}
\def\tb{\ifmmode \tan\beta \else $\tan\beta$\fi}
\def\xw{\ifmmode x\sub w\else $x\sub w$\fi}
\def\ch{\ifmmode H^\pm \else $H^\pm$\fi}
\def\lum{\ifmmode {\cal L}\else ${\cal L}$\fi}
\def\inpb{\ifmmode {\rm pb}^{-1}\else ${\rm pb}^{-1}$\fi}
\def\infb{\ifmmode {\rm fb}^{-1}\else ${\rm fb}^{-1}$\fi}
\def\epem{\ifmmode e^+e^-\else $e^+e^-$\fi}
\def\ppb{\ifmmode \bar pp\else $\bar pp$\fi}

\def\bsg{\ifmmode b\rightarrow s\gamma \else $b\rightarrow
s\gamma$\fi}

\newskip\zatskip \zatskip=0pt plus0pt minus0pt
\def\matth{\mathsurround=0pt}
\def\ltap{\ \raisebox{-.4ex}{\rlap{$\sim$}}
\raisebox{.4ex}{$<$}\ }
\def\gtap{\ \raisebox{-.4ex}{\rlap{$\sim$}}
\raisebox{.4ex}{$>$}\ }
\def\atversim#1#2{\lower0.7ex\vbox{\baselineskip\zatskip\lineski
p\zatskip
  \lineskiplimit
0pt\ialign{$\matth#1\hfil##\hfil$\crcr#2\crcr\sim\crcr}}}

\renewcommand{\thefootnote}{\fnsymbol{footnote}}

\begin{document} \begin{titlepage}
\setcounter{page}{1}
\thispagestyle{empty}
\rightline{\vbox{\halign{&#\hfil\cr
&SLAC-PUB-6580\cr
&July 1994\cr
&T/E\cr}}}
\vspace{0.8in}
\begin{center}

{\Large\bf
Constraining Anomalous Top Quark Couplings at the Tevatron}
\footnote{Work supported by the Department of
Energy, contract DE-AC03-76SF00515.}
\medskip

\normalsize D. ATWOOD, A. KAGAN, and T. G. RIZZO
\\ \smallskip
{\it {Stanford Linear Accelerator Center\\Stanford University,
Stanford, CA 94309}}\\

\end{center}

\begin{abstract}

We explore the influence of an anomalous chromomagnetic moment,
$\kappa$, onthe production characteristics of top quark pairs at the
Tevatron. We find that for top quarks in the 170 GeV mass range, present
searches are probing
values of $\kappa$ of order ${1 \over 3}$. For $\kappa$'s in this range we
find that
significant enhancements in the both the $q \bar q,~gg \to t \bar t$
production cross sections are obtained. Once top has been verified and QCD
uncertainties are under control, future high statistics measurements at the
Tevatron will eventually be sensitive to values of $\kappa$ with magnitudes
smaller than 0.10-0.15.  We discuss a class of scalar technicolor models which
may produce large values of $\kappa$ in conjunction with generation of $m_t$.

\end{abstract}

\vskip0.45in
\begin{center}

Submitted to Physical Review {\bf D}.

\end{center}


\renewcommand{\thefootnote}{\arabic{footnote}} \end{titlepage}


The possible discovery of the top quark at the Tevatron by the CDF
Collaboration{\cite {cdf,d0}} in the mass range anticipated by
precision
electroweak data{\cite {moriond}} represents a great triumph for the
Standard
Model(SM).  Once confirmed, a detailed study of the nature of the top
(\eg,
width, couplings, production properties) at both hadron{\cite {htop}}
and
\epem {\cite {etop}} colliders may yield significant information on
new
physics which lies somewhere beyond current energy scales. Existing
indirect
constraints on several of the top's properties from low energy data
are
relatively poor{\cite {low}} and leave plenty of room for new
physics.
At the present time the CDF and D0 results seem to be {\it roughly} in accord
with the expectations of QCD{\cite {qcd}}. However, the cross section as
determined by CDF does appear to be somewhat above the SM prediction which has
prompted much theoretical speculation{\cite {fermit}} as to new dynamics which
may be present.
This is shown explicitly
in Figure 1 which compares the
data from both the CDF and D0 Collaborations with the most recent
theoretical
next-to-leading order(NLO) calculations which include gluon
resummation. A well established difference between the predictions of QCD
and the Tevatron experiments would indicate the presence of new physics.
If the source of this new physics is at the TeV scale then the leading effect
should be parameterized my a QCD chromomagnetic dipole moment since this is
the lowest dimension CP-conserving effective Lagrangian contributing to the
gluon-top coupling.

In this paper we will
consider the possibility that the top quark possesses a non-zero
anomalous
chromomagnetic dipole moment, $\kappa$, in its coupling to gluons
and
explore the implications of such a scenario for top pair production at
the
Tevatron.  To get an idea of how large $\kappa$ might be
due to new physics one notes that
if the gluon is removed from a chromomagnetic dipole moment graph one is
often left with a finite contribution to the top mass.
If this is the origin of the top mass
dimensional analysis
implies that $\kappa$ is ${\cal O}\left({m_t^2 \over
\Lambda^2}\right)$, where $\Lambda$ is the scale of new physics.
As we'll see, this suggests that there is new physics below a TeV
if there is a substantial increase in the $t \bar t$
production cross-section due to non-zero $\kappa$.
Chromomagnetic dipole moments in association with quark mass
generation
can occur quite naturally in
composite models and in technicolor models.
Following a phenomenological discussion we will describe a class
of scalar technicolor models{\cite{setc,dI}} in which it may
be possible to obtain
$\kappa \sim {1 \over 4}$, resulting in an ${\cal O} (50\%)$ increase in the
$t \bar t$ production cross section at the Tevatron.

To begin our analysis, we consider the piece of the Lagrangian which
governs
the $t\bar t g$ coupling:
\begin{equation}
{\cal L}=g_s\bar t T_a \left( \gamma_\mu+i{F_2\over
{2m_t}}\sigma_{\mu\nu}
q^\nu\right)t G_a^\mu \,,
\end{equation}
where $g_s$ and $T_a$ are the usual $SU(3)_c$ coupling and
generators, $m_t$
is the top quark mass, $q$ is the gluon momentum, and $F_2$
represents a
$q^2-$dependent form factor. For $|q^2|<<\Lambda^2$, the intrinsic
scale
in the form
factor, we define $F_2=\kappa$ following the usual notation. In order
to
examine the effects of non-zero $\kappa$ on $t \bar t$ production,
we must
first calculate the parton-level $q\bar q \to t\bar t$ and $gg \to
t\bar t$
differential cross sections. For the $q\bar q$ case we obtain{\cite
{old}}
\begin{equation}
{d\sigma_{q\bar q}\over {d\hat t}}={2\pi\alpha_s^2\over {27\hat
s^2}}\left[
\left(1+{2m_t^2\over {\hat s}}\right)+3F_2+F_2^2\left({\hat s\over
{8m_t^2}}
+1\right)+{1\over {4}}(3z^2-1)\left(1-{\hat s\over
{4m_t^2}}F_2^2\right)
\right]  \,,
\end{equation}
with $\hat s$ being the parton level center of mass energy and $z$
being the
cosine of the corresponding scattering angle, $\theta^*$, as defined
via the
usual relations
\begin{eqnarray}
{\hat t} & = & {-\hat s(1-\beta z)\over {2}}+m_t^2 \,, \nonumber \\
{\hat u} & = & {-\hat s(1+\beta z)\over {2}}+m_t^2 \,,
\end {eqnarray}
with $\beta=(1-4m_t^2/{\hat s})^{1/2}$. In this expression, $F_2$ is
evaluated
at $q^2=\hat s$; note the quadratic dependence on $F_2$.
Since
top quark pair production at Tevatron for masses near 170 GeV is
dominated by
the threshold region of the $q\bar q$ annihilation process(at least for
$\kappa=1$), a brief discussion of the influence of finite $F_2$ on the
parton
level process is relevant. For $\hat s \simeq 4m_t^2$, we see that
$F_2$ has
two important effects on the differential cross section: ($i$) the
angular
dependence is softened and ($ii$) the total cross section has a
minimum at
$F_2=-1/2$ and grows rapidly as $F_2$ increases in a positive
manner away from
zero. For example, the near-threshold cross section for $F_2=0.5$ is
2.5 times
larger than for $F_2=0$. We expect these qualitative results to be
maintained
even after folding with the parton distributions and all integrations
are
performed as will be verified by explicit calculation below.

In ordinary LO and NLO QCD, for top quarks in the mass range of
interest, one
finds that the $q\bar q \to t\bar t$ subprocess contributes almost
$90\%$ of
the entire cross section. As we will see below, the dominance of this
process
remains even when $\kappa$ is non-zero provided it's magnitude is
not too
large, say, $\kappa<1$. It is thus worth while to briefly explore the
influence of $F_2$ with $\Lambda$ finite on
the LO $q\bar q \to t\bar t$ subprocess. To do this, we simply fold
the
above differential distribution with the structure functions of the
CTEQ
Collaboration{\cite {cteq}} thus obtaining the results presented in
Figure 2.
We assume for these results that $F_2$ can be simply expressed in
the simple
form
\begin{eqnarray}
F_2 & = & \kappa(1+{\hat s\over {\Lambda^2}})^{-1} \,,
\end {eqnarray}
at least as a first approximation. We see immediately that once
$\Lambda$
approaches 1-2 TeV there is not much influence from $\Lambda$
being finite
as expected from the discussion above. The reason for this is the fact
that
most of the weight of the subprocess cross section comes from the
threshold region. For simplicity, we will take $\Lambda$ to infinity in
our phenomenological calculations below.
The sensitivity of the
LO $q\bar q \to t\bar t$ subprocess to finite $\kappa$ is clearly
demonstrated
by this figure as we see that the cross section scales by factors of
order
5-10 as $\kappa$ varies between 1 and -1. This sensitivity will
persist in
the more detailed calculations below.

The case of the $gg \to t\bar t$ differential cross section is much
more
complicated; let us for simplicity consider the limit where $\Lambda>>\hat s$
so
that we can
make the replacement $F_2 \to\kappa$. Even in this `simpler'
situation, we must add an additional, dimension-5, four-point $t\bar t gg$
interaction proportional to $\kappa$ to maintain gauge
invariance{\cite {david}}. Defining the kinematic abbreviations
\begin{eqnarray}
x &=& {m_t^2\over {\hat s}} \,, \nonumber \\
K &=& {\kappa\over {2\sqrt {x}}} \,, \\
d &=& 1-z^2+4xz^2 \,, \nonumber
\end{eqnarray}
the resulting differential cross section can be written as
\begin{equation}
{d\sigma_{gg}\over {d\hat t}}={\pi\alpha_s^2\over {64\hat
s^2}}\left[
T_0+T_1K+T_2K^2+T_3K^3+T_4K^4\right]  \,,
\end{equation}
a quartic polynomial in $\kappa$, where the $T_i$ can be written as
\begin{eqnarray}
T_0 &=& 4(36xz^2-7-9z^2)(z^4-8xz^4+16x^2z^4-32x^2z^2+8xz^2-8x-1)
/{3d^2} \,, \nonumber \\
T_1 &=& -32(36xz^2-7-9z^2)\sqrt {x}/{3d} \,, \nonumber \\
T_2 &=& -16(72x^2z^2-46xz^2+7z^2-16x-7)/{3d} \,, \\
T_3 &=& 32(-7z^2+28xz^2-5x+7)\sqrt{x}/{3d} \,, \nonumber \\
T_4 &=& 16(-8xz^4+16x^2z^4+z^4-4x^2z^2+9xz^2-2z^2+1-x+4x^2)/{3d}
\,. \nonumber
\end{eqnarray}
This result is easily seen to reduce to the more conventional one
when
$\kappa \to 0$. One might expect that the sensitivity of the $gg \to
t\bar t$
differential cross section may be somewhat greater than the $q\bar
q$ case
since it is a quartic function of $\kappa$. As in the $q\bar q$ case,
near
threshold the $gg \to t\bar t$ cross section increases as $\kappa$
increases
in the positive direction. For $\kappa=0.5$, the cross section is more
than
twice as large as what one finds for $\kappa=0$.
When finite $\Lambda$ corrections become important the
calculation of the $gg \to t\bar t$ cross section becomes even more
intricate
since the form factors would then be
evaluated at $q^2=0$ in the $\hat t-$ and $\hat u-$exchange
diagrams but at
$q^2=\hat s$ in both the s-channel and four-point diagrams. The fact
that
different scales are involved results in a further violation of
$SU(3)_c$ gauge
invariance because delicate gauge cancellation are no longer taking
place.
To cure this new problem we need to add an {\it additional} four-
point
interaction, as was discussed in Ref.{\cite {david}}, whose
contribution
to the amplitude is proportional to the difference $F_2(\hat s)-
F_2(0)$.
Since the cross section is dominated by the threshold region and we are
working in the large $\Lambda$ limit, these additional contributions to
the $gg \to t\bar t$
amplitude can be ignored. Indeed, since the $gg$ contributions to
$t\bar t$
production remain sub-leading in comparison to those from $q\bar
q$ for top
masses in the 170 GeV range and values of $\kappa$ of interest to
us, we will
set $F_2(\hat s)=F_2(0)=\kappa$ in the $gg$ contribution in what
follows.

To proceed further, we follow Ref.{\cite {qcd}} and include NLO and
gluon
resummation corrections; note that these are the {\it conventional}
QCD
corrections and not the additional $\kappa$-dependent ones that can
arise in
higher order. Our philosophy will be to treat the new $\kappa$-
dependent terms
in LO only and include just the SM NLO corrections in the analysis
below.
Putting this all together we arrive at Figure 3 which shows the
separate
contributions of the $q\bar q \to t\bar t$ and $gg \to t\bar t$
subprocesses
as well as their sum in comparison the both the CDF and D0 results as
a
function of $\kappa$. Here we see explicitly some of the general
features
discussed above: ($i$) For $\kappa>(<)0$, the cross section is
larger(smaller)
than the
SM prediction; ($ii$) for $\kappa \neq 0$, the relative weights of the
$gg$
and $q\bar q$ subprocesses are altered although $q\bar q$ remains
dominant for
$\kappa>0$. For $-1\leq \kappa \leq -0.5$ we see that both
contributions are
small and have comparable magnitudes. ($iii$) To increase the total
cross section to
near the result found by CDF (and still be consistent with the D0
bound) would
require values of $\kappa$ in the approximate range $1/4-1/3$.
Certainly,
negative values
of $\kappa$ are not favored by the existing data; clearly, new top
production
cross section determinations from both the CDF and D0 collaborations
are
eagerly awaited.

If the top cross section eventually settles down to its SM value, we
can use
the results in Fig.~3 to place limits on the value of $\kappa$. Of
course,
there are many sources of both theoretical and experimental error
which play
important roles in determining the resulting allowed $\kappa$ range.
On the
theoretical side, one has to deal with (a) scale ambiguities, (b)
variations
in parton densities, and (c) NNLO corrections; the size of these
uncertainties
we can estimate from the literature. Laenen \etal {\cite {qcd}}
provide us with
an estimate of the uncertainty in the total cross section due to
various scale
choices: $+14.5\%, -8.6\%$, for tops in the mass range of interest. In a
recent paper, Martin, Stirling and Roberts(MRS){\cite {mrs}} have
discussed
the variation in the $t\bar t$ production cross section at the
Tevatron with
the choice of (modern) parton densities(PD). From their analysis, and
the fact
that top pairs are dominantly produced at large $x$, we see that the
PD
uncertainty is rather small with the variations in the central value of
the
top cross section being of order $2-3 \%$. In order to estimate the
potential
size of the NNLO corrections, we compare the MRS NLO result with
that given by
Laenen \etal , which includes gluon resummation for the same choice
of PD.
This yields an additional $4\%$ uncertainty to the cross section. If we
combine these theoretical errors with the overall scale error due to
the
Tevatron luminosity as determined by CDF{\cite {cdf}} of $3.6\%$, we
arrive at
a total uncertainty of $+15.6\%,-10.3\%$. To get a quasi-estimate of
the
experimental uncertainty, we assume that all of the error, from both
statistics and systematics(apart from the
luminosity), scales with the increase in statistics; this yields an error
of
$(+43.3,-34.3){\sqrt {19.3/{\cal L}}}$, with ${\cal L}$ being the
integrated
luminosity in $pb^{-1}$. Combining all errors in quadrature leads to
the
following estimates of the total error for ${\cal
L}=100(250,~500,~1000)
pb^{-1}$ of $(+24.6,-18.3),~(19.7,-14.0),~(17.8,-12.3),~(16.7,-11.4)$,
respectively. At $95\%$CL, these errors yield the following allowed
ranges
for $\kappa$ for the above integrated luminosities:
$-0.14\leq \kappa \leq 0.15$, $-0.11\leq \kappa \leq 0.12$,
$-0.09\leq \kappa \leq 0.11$, and $-0.08\leq \kappa \leq 0.11$,
respectively.
These results should be considered indicative of what may
eventually be
possible at the Tevatron.

Apart from the total $t\bar t$ production cross section, various
distributions
involving the top may show some sensitivity to finite $\kappa$. In
Fig. 4. we
show the $p_t$, rapidity($y$), and $t\bar t$ invariant mass($M$)
distributions
for different values of $\kappa$. As a first approximation, we see
that the
dominant effect of finite $\kappa$, especially in the case where
$\kappa$ is
positive, is to apply an approximate rescaling of the SM result by
the ratio of total cross sections. Although this might appear at first
surprising, it is merely a reflection of the fact that most of the $t\bar
t$
cross section arises from $\hat s$ values not far above threshold. Of
course,
at the highest values of $p_t$ or $M$, one begins to see small
deviations from
this simple qualitative picture, especially for values of $\kappa$ far
from
zero, but the cross sections in those parameter space regions are
always
very small. For example, the ratio of the $p_t$ distribution for
$\kappa=1$
and the SM case is approximately flat for $p_t$'s less than about 300
GeV.
However, as the $p_t$ is further increased, this ratio rises significant,
\ie ,
there is a high $p_t$ tail induced by finite $\kappa$. Of course the
cross
sections for $p_t$'s $>300$ GeV are quite small and the $\kappa=1$
case is an
extreme example. For $\kappa$'s in the $0-0.5$ range, there is very
little
sensitivity to increased $\kappa$ values in the distributions apart
from the
overall rescaling factor.

As emphasized above, the dominant effect of non-zero $\kappa$ in
the threshold
region is a simply an approximate rescaling of the SM cross section.
Of
course, near
particular values of $\kappa$ this approximation breaks down; a
special
example of this situation for the $q\bar q \to t\bar t$ subprocess,
$\kappa=-1$, can be seen immediately from Eq. (2).
For all $\kappa \neq -1$, the expression in the square brackets in Eq.
(2) is
finite whereas it vanishes for that particular value. Amongst other
things,
this would imply that the $t\bar t$ center of mass scattering
angle($z$)
distribution should be quite different when $\kappa=-1$ from all
other cases.
This expectation is borne out by the results shown in Fig. 5, which
shows the
$z=cos \theta^*$ distribution after integration over $M$ and $y$,
summing both
the $q\bar q$ and $gg$ contributions. Here we see that in all cases
the
angular dependence is quite mild, owing to threshold dominance,
except for the
case $\kappa=-1$. From Figs. 4 and 5 it is clear that additional
information on
$\kappa$ will be difficult to obtain from distribution measurements
so that we
simply have to rely on total cross section results to constrain
$\kappa$.

In order to further motivate our analysis we briefly discuss a class of
technicolor models{\cite{setc,dI}} with non-zero $\kappa$.
Consider the gauge group $G= SU(N)_{TC} \times
SU(3)_C
\times
SU(2)_L \times U(1)_Y$, together with
the following technicolored fields: a right-handed $SU(2)_L $
doublet of technileptons
$T_R (N,1,2,0)=(U_R , D_R )^T $, two left-handed $SU(2)_L$ singlet
technileptons
$U_L (N, 1,1,1/2)$,
$D_L (N, 1,1,-1/2)$, all with charges $\pm {1\over 2}$, and a charge
$ {1 \over 6 }$ color triplet techniscalar $\omega
(\overline{N}, 3,1,1/6)$.  Transformation properties with respect to
the technicolor group, $SU(N)_{TC}$, and the standard model
gauge group have been included in parenthesis.
For the purposes of our discussion we can ignore the first two
quark families. Yukawa
couplings to the third family are given by
\begin{equation}
  {\cal L}_{Y} = \lambda_Q
 \om    \overline {  Q_{ L} } { T_ R }
+
\overline{\lambda_t} \om ^* \overline{ U_L}{ t_{R }}   +
\overline{\lambda_b}
\om ^* \overline{D_L}{ b_R }  +H.c., \label{eq:tcyukawas}
\end{equation}
where $Q_L$ is the left-handed $SU(2)_L$ doublet of quarks, and
$t_R$, $b_R$ are the right-handed $SU(2)_L$ singlet quarks.
$\om$ acquires a mass from the scalar sector of the
Lagrangian and a
`constituent' mass from technicolor dynamics.\footnote{Scalar technicolor
models
can be supersymmetrized in order to protect the masses of the scalars.
In turn, supersymmetric flavour-changing neutral currents can be suppressed
since a multi-TeV supersymmetry breaking scale is natural in this
framework{\cite{dks}}.}

Technifermion
condensates will induce top and
bottom quark masses via techniscalar exchange\footnote{Additional
quark masses can be generated by adding more techniscalars,
more technileptons, or Higgs doublets which acquire small
vacuum expectation values by coupling to the technilepton condensates.
Radiative mass contributions could, in principle, also play a role
for light quark masses.}
, in analogy
with fermion mass generation via gauge boson
exchange in extended technicolor models.
In the limit $m_{\om} >> \Lambda_{TC}$, where
$\Lambda_{TC} \sim 1$ TeV, $\om$ can be integrated out and one obtains
\begin{equation}
 m_t \approx \lambda_Q \overline{\lambda_t}
{{\langle \overline {U} U \rangle } \over {4m_{\om}^2
}},~~~~m_b \approx \lambda_Q \overline{\lambda_b}
{{\langle \overline {D} D \rangle } \over {4m_{\om}^2
}}.
\end{equation}
The magnitude of the condensates is
estimated to be{\cite{georgimanohar}}
\begin{equation} \langle \overline {D} D
\rangle = \langle \overline {U} U
\rangle
\approx \left({3\over N_{TC}}\right)^{1\over 2} 4 \pi
\left({v \over \sqrt{N_D}}\right)^3~{\rm GeV}^3 ,
\end{equation}
where
$v=246$ GeV, and $N_D$ (equal to one above) is the number of
technifermion doublets, $T_R$.
Chromomagnetic dipole moments are due to
emmision of a gluon by the exchanged techniscalar.  One obtains
\begin{equation}
  {\kappa\over {2m_t}}  \approx {{m_t (m_{\om}) }\over m_\om^2}
\end{equation}
at $m_{\om}$.
Leading-order QCD evolution from TeV scales
to $\mu \sim 2 m_t$ will reduce $\kappa$ by a few percent and
can be neglected for our purposes.

For $\kappa$ to have a substantial effect
on the $t \bar t$ production cross section
$m_{\om} $ must be small.
Unfortunately, for $m_{\om} \sim \Lambda_{TC}$
we can no longer simply integrate
$\om$ out to obtain expressions for $m_t$ and $\kappa$
since strong technicolor dynamics become important.\footnote{For example, it
may
be that
the exchanged techniscalar and technifermion
bind so that the quark's mass can be attributed to mixing
with composite heavy quarks. }  Nevertheless, we expect the above expressions
to
give the correct orders of
magnitude and we defer
a more sophisticated treatment to future investigation.
Guided by estimates of the technifermion constituent
mass{\cite{chivukula}}, obtained by scaling of the QCD constituent
mass (one obtains $m_{TC} \sim (300~{\rm MeV}){v \over {\sqrt{N_D} f_\pi }}$
or $800$ GeV for $N_D =1$, $550$ GeV for $N_D =2$), we assume that
$m_{\om} \gtap {1 \over 2}$ TeV is a reasonable range to take in eqs.
(9) and (11).
\footnote{
Quark-techniscalar Yukawa couplings can vary
substantially so that all quark masses can be generated with ${\cal O} ({1\over
2})$ TeV techniscalars. Furthermore, such light techniscalar
masses do not pose a danger for flavor-changing neutral currents since
the latter first arise at the one-loop level.}
So for $m_t \approx 170~GeV$ we expect $\kappa \ltap {1 \over 4}$.
Assuming a form factor of the form given in eq. (4), with $\Lambda$
identified with $m_{\om}$, Figs. 2 and 3
imply that ${\cal O}(50\%)$
increases in the Tevatron
$t \bar t$ production cross section may be possible.

In this paper, we have considered the influence of a non-zero
chromomagnetic
moment for the top quark, $\kappa$, on the production of $t\bar t$
pairs at
the Tevatron for top masses near 170 GeV. Non-zero values of
$\kappa$ may be
present in both compositeness and technicolor scenarios. In
particular, our results can be summarized as follows:

($i$) We have obtained born-level expressions for $q\bar q,gg \to
t\bar t$ for
arbitrary values of $\kappa$ and used the SM NLO and gluon
resummation
`K-factors' from{\cite {qcd}} to obtain total cross sections and various
distributions for top pair production at the Tevatron.

($ii$) We explored the possible influence of including form factors
with a
finite scale parameter, $\Lambda$, instead of a simple constant value
for
$\kappa$. We found, since the cross section was dominated by
$t\bar t$
invariant masses not far from threshold, that values of $\Lambda$ in
the 1-2
TeV or above were essentially indistinguishable from
$\Lambda=\infty$. However, for smaller values of $\Lambda$, the
$\kappa$ dependence was found to be softened.

($iii$) For top masses in the $m_t=170$ GeV range, we demonstrated
that the
top pair production cross section was quite sensitive to the value of
$\kappa$. Values of $\kappa$ in the range $1/4-1/3$ were shown to
increase the
SM cross section to the level reported by CDF while still remaining
consistent
with the bounds from D0.  Since $\kappa$ is
${\cal O}({m_t^2 \over{\Lambda^2}}) $ if associated
with top mass generation, such large values
would likely be
due to new physics at a scale $\Lambda$ below a TeV.
If the cross section was eventually found
to agree
with the SM expectations, we estimated the bounds on $\kappa$
obtainable at
the Tevatron as the integrated luminosity increases. Included in this
analysis
are uncertainties due to scale ambiguities, structure function
variations,
luminosity uncertainties, and estimates of NNLO contributions, as
well as
statistics. We found that from the total cross section alone the
Tevatron will
be able to probe values of $|\kappa|<0.10-0.15$ in the not too distant
future, potentially providing us with a new window to physics in the TeV
region.

($iv$) We explored the possibility that $p_t$, rapidity($y$), top pair
mass($M$), and center of mass scattering angle($cos \theta^*$)
distributions
may provide additional constraints on a potential non-zero value for
$\kappa$.
This analysis found that once these distributions were rescaled by
the ratio
of the $\kappa$-dependent to SM cross section almost all of the
sensitivity
was found to lie in parameter regions where differential cross
sections were
very small. Our conclusion is that these various
distributions are
probably
not too useful in obtaining additional constraints on $\kappa$
beyond those
obtainable from the total cross section.

($v$)  We discussed a class of scalar technicolor models in which
both the top's mass
and $\kappa \sim {1 \over 4}$
could be due to exchange of a techniscalar with ${1\over 2}$ TeV
mass.  It is interesting that with the techniscalar's mass at
${1\over  2}$ TeV (flavor-changing) chromomagnetic dipole moments can
also lead to suppression of the $B$ semileptonic branching ratio and
substantial $\Delta {\rm I}={1 \over 2}$ enhancement in
$K$ decays{\cite{dI}}.

If an anomalous chromomagnetic moment for the top quark exists it
will open a new window to new physics beyond the Standard Model.

\vskip.25in
\centerline{ACKNOWLEDGEMENTS}

TGR would like to thank
W.J. Stirling for detailed numerical results on top pair production at
the
Tevatron in NLO for various choices of the parton distributions. He
would
also like to thank L. Dixon, J.L. Hewett, N. Hadley, T. Barklow, P.
Burrows,
S. Brodsky and S. Wagner for discussions related to this work.
TGR would also
like to thank the members of the Argonne National Laboratory High
Energy
Theory Group for use of their computing facilities.

\newpage

%
\def\MPL #1 #2 #3 {Mod.~Phys.~Lett.~{\bf#1},\ #2 (#3)}
\def\NPB #1 #2 #3 {Nucl.~Phys.~{\bf#1},\ #2 (#3)}
\def\PLB #1 #2 #3 {Phys.~Lett.~{\bf#1},\ #2 (#3)}
\def\PR #1 #2 #3 {Phys.~Rep.~{\bf#1},\ #2 (#3)}
\def\PRD #1 #2 #3 {Phys.~Rev.~{\bf#1},\ #2 (#3)}
\def\PRL #1 #2 #3 {Phys.~Rev.~Lett.~{\bf#1},\ #2 (#3)}
\def\RMP #1 #2 #3 {Rev.~Mod.~Phys.~{\bf#1},\ #2 (#3)}
\def\ZP #1 #2 #3 {Z.~Phys.~{\bf#1},\ #2 (#3)}
\def\IJMP #1 #2 #3 {Int.~J.~Mod.~Phys.~{\bf#1},\ #2 (#3)}

\newpage

%
{\bf Figure Captions}
\begin{itemize}

\item[Figure 1.]{Theoretical NLO cross section(dash-dotted curve) for
$t\bar t$ production at the Tevatron, including gluon resummation,
as a
function of the top quark mass from the work of
Laenen \etal {\cite {qcd}} and the corresponding anticipated
uncertainty
due to scale choice(dotted curves). The data point is the CDF result,
while
the horizontal dashed line is the $95\%$ CL upper limit reported by
the D0
Collaboration.}
\item[Figure 2.]{LO calculation of $q\bar q \to t\bar t$ production
cross
section using CTEQ structure functions assuming $m_t=170$ GeV as a
function
of $\kappa$. The dotted(dashed, dash-dotted, solid, square-dotted)
curves
correspond to $\Lambda=0.25(0.5,~1,~2,~\infty)$ TeV, respectively.}
\item[Figure 3.]{NLO cross sections for the $q\bar q \to t\bar
t$(dash-dotted)
and $gg \to t\bar t$(dotted) subprocesses as well as the total cross
section(solid) at the Tevatron as functions of $\kappa$ for
$m_t=170$ GeV
using the CTEQ parton distribution functions. The horizontal dashed
lines
provide the $\pm 1\sigma$ CDF cross section determination while
the horizontal
dotted line is the D0 $95\%$ CL upper limit.}
\item[Figure 4.]{(a) $p_t$ distribution for top quark pairs produced
at the
Tevatron assuming $m_t=170$ GeV and CTEQ PD. The solid curve is
the SM
prediction and the upper(lower) dash-dotted, dashed, and dotted
curves
correspond to $\kappa=1,~0.5,~0.25(-1,~-0.5,~-0.25)$, respectively.
(b) Top
pair invariant mass distributions for the same cases as shown in (a).
(c) Top
quark rapidity distributions for the same cases as shown in (a).}
\item[Figure 5.]{$cos \theta^*$ distribution for top-pair production as
in
Figure 4, except that the upper(lower) dash-dotted, dashed,
and dotted curves correspond to $\kappa=0.25,~0.5,~1(-0.25,$~$~-0.5,~-
1)$,
respectively.}
\end{itemize}

\end{document}